\documentclass[twocolumn,english,aps,prb,showpacs,superscriptaddress,floats,amsmath,amssymb,floatfix]{revtex4}

\usepackage{amsmath}
\usepackage{amssymb}
\usepackage{wasysym}
\usepackage{psfrag}
\usepackage{tabularx}
\usepackage[dvips]{graphicx}
\usepackage{enumitem}
\usepackage{color}
\def\beq{\begin{equation}}
\def\eeq{\end{equation}}
\def\bea{\begin{eqnarray}}
\def\eea{\end{eqnarray}}
\def\nn{\nonumber}

\def\e{\vec{e}}

\def\trigup{\bigtriangleup}
\def\trigdown{\bigtriangledown}

\begin{document}

\title{Degenerate and chiral states in the extended Heisenberg model in the kagome lattice }
\author{F.A. G\'omez Albarrac\'in}
\affiliation{IFLP-CONICET. Facultad de Ciencias Exactas, Universidad Nacional de La Plata, C.C. 67, 1900 La Plata, Argentina and Departamento de F\'{i}sica, FCE, UNLP, La Plata, Argentina}
\affiliation{Laboratoire de Physique Th\'eorique, IRSAMC, CNRS and Universit\'e de Toulouse, UPS, F-31062 Toulouse,France}
\author{P. Pujol}
\affiliation{Laboratoire de Physique Th\'eorique, IRSAMC, CNRS and Universit\'e de Toulouse, UPS, F-31062 Toulouse,France}
\begin{abstract}
We present a study of the low temperature phases of the antiferromagnetic extended classical Heisenberg model in the kagome lattice, up to third nearest 
neighbors. First, we focus on the degenerate lines in the boundaries of the well-known staggered  chiral phases. These boundaries have either semi-extensive
or extensive degeneracy, and we discuss the partial selection of states by thermal fluctuations. Then, we study the model under an external magnetic field on these 
lines and in the staggered chiral phases. We pay particular attention to the highly frustrated point, where the three exchange couplings are equal. We show that this point 
can me mapped to a model with spin liquid behavior and non-zero chirality. Finally, we explore the effect of Dzyaloshinskii-Moriya (DM) interactions in two ways:
an homogeneous and a staggered DM interaction. In both cases, 
there is a rich low temperature phase diagram, with different spontaneously broken symmetries and non trivial  chiral phases.
\end{abstract}

\maketitle

\section{Introduction}
In the last years, much effort has been devoted to the study of topological phases in magnetic systems.
 Experimental results in Herbersmithite and related materials, where the magnetic ions form a kagome lattice, have
been one of the main motivations for the large amount of theoretical work in the kagome lattice.
Different models have been proposed and explored, in the quest for exotic topological phases (for a recent review on 
experimental and theoretical results, see Ref.[\onlinecite{ReviewNorman}] and references therein.). Much of these works
have been done in 
the so-called extended Heisenberg model,
including exchange interactions up to third nearest neighbors. Classically, for a certain region in parameter space, the ground-states of this model present 
a non-coplanar order with staggered chirality \cite{Domenge12subl,Richter12subl,MessioCuboc,MessioPhases}. In the quantum case, 
recent studies  focused in ferromagnetic first
and second nearest neighbors couplings and a dominant antiferromagnetic third nearest neighbors coupling. They proposed the existence of chiral spin liquids
and different phases such as valence bond crystals for these models
\cite{ShengBalents1,MessioCSL}. For antiferromagnetic couplings, the DMRG study presented in Ref.[\onlinecite{ShengSciRep}] states that along
the boundary between a  staggered chirality and a $q=0$ phase in the classical phase diagram,
there is a quantum chiral spin liquid phase with fractional quantum Hall effect excitations. 
This boundary and the nature of the quantum phases was further studied in Refs.[\onlinecite{ShengBalents2}].

In this work we focus on the frustrated extended antiferromagnetic model in the kagome lattice for classical spins, in the phases
with staggered  chirality  and its boundaires. 
Our main goal is to study the low temperature behavior of these regions under the effect of different additional interactions, 
looking for non-trivial chiral phases. In order to do this, we first discuss the degenerate lines of the isotropic model and the effect of a magnetic 
field. These lines have different types of semi-extensive and extensive degeneracy. Most interestingly, they intersect in a highly frustrated point
which can me mapped to a model with net chirality and spin liquid characteristics. Then, we enhance chiral phases by adding two types of antisymmetric
Dzyaloshinskii-Moriya interaction: an homogeneous one, which has been widely studied for the kagome antiferromagnet, and a staggered one. 
We find that in the classically chiral region these interactions give rise to a series of non trivial low temperature phases, with
spontaneously broken symmetries and different types of non-zero chirality.

The manuscript is organized as follows. In Sec. II we present the model, indicating the different terms considered in the Hamiltonian. We briefly discuss 
the symmetries
associated with each term, and comment on the way the Monte Carlo simulations were implemented. Then, in the next sections we focus on the low
temperature phases of the 
isotropic model (Sec.III), the effect of an external magnetic field (Sec.IV) and of Dzyaloshinskii-Moriya interactions (Sec.V). We conclude in Sec.VI
with an overall discussion and remarks for future work.

\section{The model}

\begin{figure}[h!]
\includegraphics[width=0.8\columnwidth]{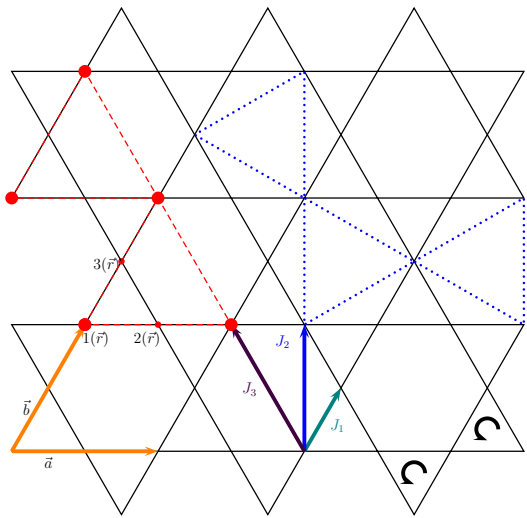} 
\caption{\label{fig:latt} Kagome lattice. Three sites of a unit cell ($1,2,3$), lattice vectors $\vec{a}, \vec{b}$ and 
first, second and third nearest neighbors couplings $J_1,J_2,J_3$
for site type 1 are indicated. Black solid arrows indicate the anticlockwise direction in which the bonds are taken for the Dzyaloshinskii-Moriya
interaction. Dashed red lines form a portion of the site 1 triangular sublattice
and blue dotted lines indicate part of one of the three interpenetrating kagome sublattices 
formed by second nearest neighbor bonds.}
\end{figure}

In this section we describe the different terms of the Hamiltonian of the model defined in the kagome lattice for classical spins, the different relevant symmetries and
the  Monte Carlo technique that was used to explore the low temperature phases.
In Fig. \ref{fig:latt} we show the kagome lattice and indicate the lattice vectors $\vec{a}=(1,0)$ and $\vec{b}=(1/2,\sqrt{3}/2)$,
the three sites per unit cell $\vec{r}$, and the bonds and sublattices considered in this work.

\subsection{Hamiltonian}
The Hamiltonian of the system in consideration contains three different parts:

\beq
\mathcal{H}= \mathcal{H}_I + \mathcal{H}_M + \mathcal{H}_{DM}
\label{FullH}
\eeq
with the first term corresponding to the fully isotropic Heisenberg Hamiltonian:
\bea
\mathcal{H}_I = J_1 \sum_{\langle i,j \rangle}\vec{S}_i\cdot\vec{S}_j  + J_2 \sum_{\langle \langle k,l  \rangle \rangle}\vec{S}_k\cdot\vec{S}_l 
+ J_3 \sum_{\langle \langle \langle m,n \rangle \rangle \rangle}\vec{S}_m\cdot\vec{S}_n 
\eea
where $\langle  .,. \rangle$, $\langle \langle .,.  \rangle \rangle$ and $\langle \langle \langle .,. \rangle \rangle \rangle$ mean respectively first, second 
and third neighbors as shown in Fig. \ref{fig:latt}. Note that we take only the third nearest neighbor bonds that cross the hexagons.
The second term corresponds to the presence of a magnetic field, chosen along the $z$ direction:
\beq
 \mathcal{H}_M = -h \sum_i S_i^z
\eeq
while the last term includes an out-of-plane Dzyaloshinskii-Moriya (DM) interaction, in the same direction as the external magnetic field:
\beq \label{eq:HDM}
\mathcal{H}_{DM} = \vec{D}_\trigup\cdot\sum_{\langle i,j \rangle_\trigup}\vec{S}_i\times\vec{S}_j + 
\vec{D}_\trigdown\cdot\sum_{\langle i,j \rangle_\trigdown}\vec{S}_i\times\vec{S}_j 
\eeq
where $\langle i,j \rangle$ denote nearest neighbors sites in up and down triangles in the kagome lattice 
($\trigup$ and $\trigdown$) taken in the trigonometric sense order shown in Fig. \ref{fig:latt}.

\subsection{Relevant symmetries}
We briefly discuss the symmetries of the above described Hamiltonian relevant to this work.
The presence of a magnetic field in the interaction Hamiltonian 
\beq \label{eq:HIM}
\mathcal{H}= \mathcal{H}_I + \mathcal{H}_M 
\eeq
\noindent reduces the  $SO(3)$ rotation symmetry group from $\mathcal{H}_I$  to an $SO(2)$, 
which allows us to expect a Berezinsky-Kosterlitz-Thouless (BKT) algebraic phase at low temperatures. 
Time Reversal Symmetry (TRS) is also broken because of the magnetic field but inversion and reflection symmetries are still present.

In this work, we have studied the DM term in Eq.(\ref{eq:HDM}) in two different ways: considering  a uniform (Eq.(\ref{eq:DMeq})) 
or a staggered (Eq.(\ref{eq:DMdiff})) $\vec{D}$:
\bea
\vec{D}_\trigup = \vec{D}_\trigdown = D\e_z \label{eq:DMeq} \\
\vec{D}_\trigup = D \e_z ~;~ \vec{D}_\trigdown = -D \e_z \label{eq:DMdiff}
\eea

The uniform case is the one that has been more widely studied in the literature \cite{CanalsLacroixDM,MikeOctu,LhuillierDM1,LhuillierDM2,Jaubert2017}. 
It preserves inversion and reflexion symmetries (keeping in mind that spins are axial vector). 
For the first neighbors kagome model it is know to favor a $q=0$ state with homogenous chirality per triangle. 

As we will discuss in section \ref{sec:HDM},
the inclusion of a staggered DM interaction induces a richer low temperature phase diagram. The staggered interaction in Eq.(\ref{eq:DMdiff})
breaks inversion and reflexion symmetries, but preserves the lattice only reflexion symmetry $(x, y) \to (x,-y)$. 
Of course in both cases the $SO(2)$ rotation symmetry around the $z$ axis is still present.

\subsection{MonteCarlo Method}

To study the low temperature phases of the system we resorted to Monte Carlo simulations combining the Metropolis algorithm with the over-relaxation
method (microcanonical updates) \cite{MC}. Simulations where done for $N=3L^2$ site lattices with periodic boundary conditions,
$L=24-48$, reaching temperature $T/J_1=2\times10^{-4}$ with the annealing
technique following a $T_{i+1}=0.9T_i$ scheme. Measurements where taken after $0.1-10\times10^{5}$ Monte Carlo steps (MCS) 
for initial system relaxation and twice as much MCS for calculating  mean values.

\section{The isotropic case} \label{sec:HI}
\begin{figure}[h!]
\includegraphics[width=0.9\columnwidth]{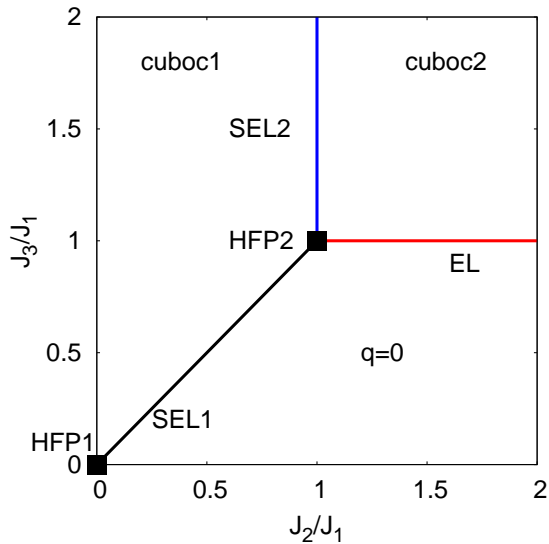} 
\caption{\label{fig:PhaseD} Ground state phase diagram for the antiferromagnetic $J_1-J_2-J_3$ extended Heisenberg model in the kagome lattice.
``HFP'' indicate the Highly Frustrated Points, ``SEL'' de semi extensive degeneracy lines and ``EL'' the extended degeneracy line. This phase 
diagram is taken from Ref.[\onlinecite{MessioPhaseD}]}
\end{figure}

The isotropic case corresponds to $h=D=0$ leaving only the term $\mathcal{H}_I$. 
The ground state structure was studied in Ref.[\onlinecite{MessioPhases}] and shows a rich phase diagram depicted in Fig. \ref{fig:PhaseD}.
This phase diagram has three different phases: a $q=0$ one and two  staggered chiral phases. We name them 
 {\it cuboc-1} and {\it cuboc-2},  following the notation in the works of Messio et al. \cite{MessioPhases} 
 where ``cuboc'' stands for ``cuboctahedron''.
In these phases, in the ground state the magnetic unit cell is formed by 12 spins,  pointing at each of 
the 12  vertexes of a cuboctahedron.
In the {\it cuboc-1} state, the nearest-neighbor spins form  a $2\pi/3$ angle, and in {\it cuboc-2} they form an angle of $\pi/6$. This implies that in 
{\it cuboc-1} there is a staggered chirality in the second nearest neighbor triangles, whereas in {\it cuboc-2} it is in the first nearest neighbor ones.
Note that along the $J_2=J_3$ line, the Hamiltonian can be rewritten as in Eq.(\ref{eq:Hsle1}). Inside the {\it cuboc-2} phase, this implies that the lowest energy configuration requires the maximization of the total spin of the up and down corner-sharing
triangles and the minimization of the total spin in the hexagons.

In Ref.[\onlinecite{MessioCuboc}] it was shown that in the {\it cuboc-2} phase, 
the system undergoes a phase transition from a high temperature trivial paramagnet to a staggered chiral phase, 
in which the rotation symmetry is of course restored by fluctuations but the discrete TRS and lattice reflection
symmetries  are spontaneously broken. 

In this work, we define scalar chirality in the up and down triangles of the kagome lattice as
\begin{eqnarray} \label{eq:Chir}
\chi_{\trigup}=\vec{S}_1(\vec{r})\cdot\left(\vec{S}_2(\vec{r}) \times \vec{S}_3(\vec{r})\right) \\
\chi_{\trigdown}=\vec{S}_1(\vec{r})\cdot\left(\vec{S}_3(\vec{r}-\vec{b}) \times \vec{S}_2(\vec{r}-\vec{a})\right)
\end{eqnarray}

We now focus on the low temperature physics of the degenerate lines, described below.

As a first approach to study the degeneracy of the system along these lines, we use the spherical approximation or $O(N)$ model, where
the Hamiltonian matrix in Fourier space is diagonalised relaxing  the local $|\vec{S}_i|=1$ constraint to a global $|\sum_i^N\vec{S}_i|=N$. 
As this is a non-Bravais lattice,
this method provides a lower bound to the ground-state energy, and not the exact ground state energy. However, it  does provide useful information
about the characteristics and physics of the ground states (for a discussion on this approach  see for example Ref.[\onlinecite{Vish}]).

Of particular interest in the phase diagram are the points dubbed HFP1 (Highly Frustraded Point),
which is the well known first nearest neighbors model, and corresponds to corner sharing triangles, and HFP2, 
corresponding to corner sharing hexagons, studied in Ref.[\onlinecite{MoessnerkagomeHFP2}].
Although apparently similar, the low temperature physics of these two maximally frustrated points in the kagome lattice is
quite different. For HFP1 it has been shown that, by an Order-By-Disorder (OBD) mechanism \cite{OBDVillain}, thermal fluctuations
select coplanar states which have quartic soft modes and thus a lower contribution to the specific heat than $C_v=1$ ($k_B \equiv 1$)
predicted by the equipartition 
theorem. Therefore at low temperatures,  the specific heat of the system  tends to the well known value of $C_v={11/12}$ 
\cite{kagomeOBD}. On the other hand, HFP2 has recently been established as  a non algebraic classical spin liquid \cite{MoessnerkagomeHFP2}. Evidence of these
difference between HFP1 and HFP2 is reflected in the spherical approximation analysis: at both points the lowest band is flat,
but in HFP2 it is doubly degenerate and separated by a gap from non-flat bands, a signature of the non-algebraic nature of the spin liquid phase\cite{MoessnerkagomeHFP2}.

\begin{figure}[h!]
\includegraphics[width=1.0\columnwidth]{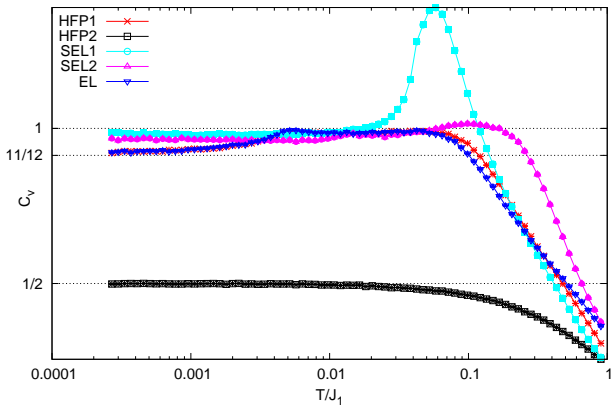}
\caption{\label{fig:CvT} Specific heat per spin ($k_B\equiv1$) as a function of temperature for different degenerate points in the kagome lattice. Results shown
correspond to the average of 10 copies of $L=30$ simulations.}
\end{figure}

An interesting quantity to estimate the 
degree of degeneracy is given by the number of  degrees of freedom per plaquette $F$ (see for example Ref.[\onlinecite{MoessnerHoneyJ3}]):

\beq \label{eq:Fplaq}
F=\frac{q}{b}\left(n-1\right)-n 
\eeq
\noindent where $n$ is the number of components of the spin, $q$ the number of spins per plaquette and $b$ the number of plaquettes that share the same spin.
For example, in HFP1 this quantity is know to be 0. On the other hand, for HFP2, one obtains $F=3$ 
($q=6$ and $b=2$), a significantly big number compared with other classical spin liquids
such as the antiferromagnetic
model in the pyrochlore lattice \cite{MoessnerPiro}, where $F=1$. Another interesting quantity to look at is the value of the specific heat at low temperature.
Each quadratic mode contributes to it with a value of $1/2$ (in units of $k_B$), each quartic mode with $1/4$, and there is no contribution from zero modes.
A first indication to check whether there is selection due to OBD is to look at the value of the specific heat: if there is no selection by an extensive number of soft modes, one expects the system to lie in a typical ground state with a non extensive number of soft modes.
Thus, the decrease of the specific heat is due only to zero modes and  reaches the value $C_v=bnk_B/2q$\cite{MoessnerPiro}. Let us come back to the case of HFP2.
We have measured a low temperature value of $C_v=1/2$ (Fig. \ref{fig:CvT}), which is the first condition to fulfil
for a disordered state.
\footnote{Note that the calculation of the degrees of freedom per plaquette (Eq.(\ref{eq:Fplaq}))
for $XY$ spins ($n=2$) gives $F=1$, and a specific heat of $C_v=1/3$.
We have checked this and the structure factors with Monte Carlo simulations, 
showing that for the HFP2 there does not seems to be thermal selection, indicating that the $XY$ case shows also a spin liquid behavior.
This implies that this system is expected to be an XY model with no Berezinsky-Kosterlitz-Thouless transition.}.

Recently, in Ref.[\onlinecite{JaubertNatComm}] it was shown that through  local transformations it is possible to do a one-to-one mapping
from the HFP1 point to two models with finite vector chirality, dubbed models $\chi^{+}$ and $\chi^{-}$. This mapping consists on a rotation of each of the three site sublattices around the $z$ axis.
The resulting Hamiltonians are an XXZ model with ferromagnetic interactions in the $xy$ plane and antiferromagnetic ones in $z$, and a 
Dzyaloshinskii-Moriya term with $D=\pm\sqrt{3}J/2$:

\beq \label{eq:hfp1xxz}
\mathcal{H}^{HFP1}_{\chi^{\pm}}=\frac{J}{2}\sum_{\langle i,j\rangle}-\vec{S}_i^{\perp}\cdot\vec{S}_j^{\perp}+2S_i^zS_j^z \pm \sqrt{3}\tilde{z}\cdot\left(\vec{S}_i\times\vec{S}_j\right)
\eeq

As mentioned in Ref.[\onlinecite{JaubertNatComm}], the low temperature physics of the $\chi^{+}$ and $\chi^{-}$ models is equivalent to the one of the HFP1. The system
undergoes an order-by-disorder state selection and the specific heat tends to $11/12$.

These same transformations can be applied to the case of second and third neighbors coupling that we are considering here. It is particularly interesting when applied to the HFP2. 
As with the first nearest neighbor bonds, the second nearest neighbors 
are between spins of different sublattices. Therefore, the transformations affect both the $J_1$ and the $J_2$ terms in the same way. However,
the third nearest neighbor couplings are between sites of the same sublattices, and thus are unchanged by this rotations. The resulting 
Hamiltonians are:

\bea \label{eq:hfp2xxz}
\mathcal{H}^{HFP2}_{\chi^{\pm}}=\frac{J_1}{2}\sum_{\langle i,j\rangle}-\vec{S}_i^{\perp}\cdot\vec{S}_j^{\perp}+2S_i^zS_j^z \pm \sqrt{3}\tilde{z}\cdot\left(\vec{S}_i\times\vec{S}_j\right) \nn \\
+\frac{J_2}{2}\sum_{\langle\langle k,l\rangle\rangle}-\vec{S}_k^{\perp}\cdot\vec{S}_l^{\perp}+2S_k^zS_l^z \pm \sqrt{3}\tilde{z}\cdot\left(\vec{S}_k\times\vec{S}_l\right) \nn\\
+J_3\sum_{\langle\langle\langle m,n \rangle\rangle\rangle} \vec{S}_m\cdot\vec{S}_n \,\,\,\,\,\,\,\
\eea

Given that the HFP2 is a spin liquid at low temperatures, so are the models in Eq.(\ref{eq:hfp2xxz}). We have indeed checked that they have the same specific heat at low temperature.
The interplay between the antisymmetric DM terms and 
an external magnetic field has interesting consequences for these models, namely, a classical chiral spin liquid, as we will discuss in the following subsection.

\begin{figure}
\includegraphics[width=0.9\columnwidth]{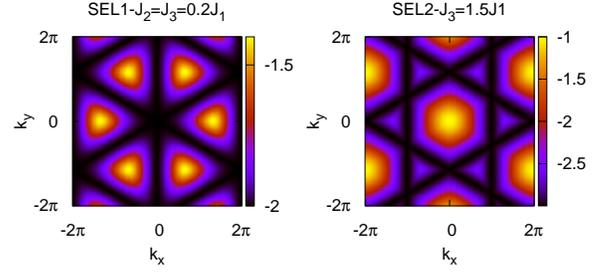}
\caption{\label{fig:SpherApp} Lowest energy bands obtained with the spherical approximation for the semi-extensive degeneracy lines SLE1 (left) and SLE2 (right)
of the phase diagram in Fig. \ref{fig:PhaseD} from the Hamiltonian presented in Eq.(\ref{eq:HIM}). }
\end{figure}

We now proceed to comment on the degenerate lines of the phase diagram.
The lines SEL1 and SEL2 are characterised by a lowest energy configuration with semi-extensive degeneracy: the submanifold of lowest energy states corresponds to lines instead of planes, as can be seen from the results of the spherical approximation
for two arbitrary points in these lines in Fig. \ref{fig:SpherApp}.

  Along the  $J_2=J_3$ line, the Hamiltonian can be rewritten as a combination of corner sharing triangles and hexagons:
  
\begin{equation} \label{eq:Hsle1}
\mathcal{H}_{J_2=J_3} = \frac{(J_1-J_2)}{2}\sum_{\trigup,\trigdown}\vec{S}_{\trigup,\trigdown}^2 + \frac{J_2}{2}\sum_{\hexagon}\vec{S}_{\hexagon}^2 
\end{equation}

From the Hamiltonian in Eq.(\ref{eq:Hsle1}) it is clear that in the SLE1 line, $J_2=J_3<J_1$, the ground states is obtained by the simultaneous minimisation of all hexagon and triangle variables (which can be shown to be independent variables).
The result corresponds to simultaneously impose zero total magnetisation in both
types of plaquettes. In this case, thermal fluctuations select those ground states with semi extensive degeneracy. Here, degenerate configurations can be build by the exchange of lines of spins
in two of the three sublattices of the kagome lattice. In the SLE2 line ($J_1=J_2<J_3$), the degeneracy line is realized in states where in each sublattice 
lines of  antiferromagnetically ordered third nearest neighbors can be independently rotated without energy cost. 
The specific heat, in the thermodynaminc limit,
for both SLE1 and SLE2 will thus tend to 1  at low temperatures (Fig. \ref{fig:CvT}), since there are no soft modes and only a semiextensive number of zero modes.

Instead, the line EL $J_1=J_3<J_2$ is a fully extensive degeneracy line,
as indeed here the Hamiltonian can be written as three intercalated kagome lattices constructed by second nearest neighbor spins (see Fig. \ref{fig:latt}):
\bea
\mathcal{H}_{J_1=J_3<J_2} = {(J_2 - J_1) \over 2}  \sum_{\rhd,\lhd,j=A,B,C} \left( \vec{S}_{(\rhd,\lhd)_j}^2  \right)\nn \\
+ {J_1 \over 2} \left[ \sum_{\rhd_A,\lhd_B} \left( \vec{S}_{\rhd_A}+ \vec{S}_{\lhd_B}  \right)^2 \right.
+ \sum_{\rhd_B,\lhd_C} \left( \vec{S}_{\rhd_B}+ \vec{S}_{\lhd_C}  \right)^2  \nn \\
+ \left. \sum_{\rhd_C,\lhd_A} \left( \vec{S}_{\rhd_C}+ \vec{S}_{\lhd_A}  \right)^2 \right] &\,\,\, 
\eea
where $\vec{S}_{\rhd,\lhd,j=A,B,C}$ is the total magnetisation of a triangle constructed by second nearest neighbors, in analogy to $\trigup$
and $\trigdown$ for the first nearest neighbor triangles. Minimization of the above Hamiltonian
easily renders the condition that the total magnetization per triangle be zero as in HFP1.
The same low temperature physics as HFP1 is thus expected, and we confirm this with the specific heat 
obtained from Monte Carlo simulations (Fig. \ref{fig:CvT}), where at low temperatures  $C_v \sim {11/12}$.

\section{The system in a magnetic field}

The low temperature phases of the kagome model under a magnetic field at HFP1 have been extensively discussed \cite{MikeKagH}, 
where in particular a collinear $UUD$ pseudoplateau around $M=1/3$ emerges ($M=\sum_i^NS^z_i$), from thermal order
by disorder. In the HFP2 point, no pseudoplateau is found, due to the high degeneracy of the model.

Between HFP1 and HFP2 along the SLE1 line, the semi-extensive degeneracy 
is still present and the system also shows a $M=1/3$ pseudoplateau. To understand this, one can rewrite the magnetic field terms of the Hamiltonian as

\begin{equation} \label{eq:HhzSLE1}
\mathcal{H}_{M,J_2=J_3}=-h\sum_{\hexagon,\trigup,\trigdown}\frac{2J_2}{J_1+J_2}S^z_{\hexagon} + \frac{J_1-J_2}{J_1+J_2}S^z_{\trigup,\trigdown}
\end{equation}

Minimising again the Hamitonian given by Eq.(\ref{eq:Hsle1}) and Eq.(\ref{eq:HhzSLE1}),
one finds that for the value of the magnetic field $h = J_1 + J_2$, the value of the magnetisation for both the triangles and the hexagons must be one third of its total value.
As happens for the nearest neighbors kagome model, the pseudoplateau occurs for a collinear configurations with an $UUD$ arrangement 
in the corner sharing triangles and a $UUUUDD$ one in the corner sharing hexagons.

The SLE2 line shows no particular features in the magnetization curves and the semi-extensive degeneracy remains,
whereas in the EL line the behavior is equivalent to the nearest-neighbor kagome
model, as expected. At low temperatures, a collinear pseudoplateau emerges, where the $UUD$ order is found in the  triangles 
of each of the three second nearest neighbor sublattices.

In the {\it cuboc} phases, the staggered chirality phase with broken reflection symmetries 
persists for low magnetic fields. By increasing further the magnetic field one drives the system into the trivial paramagnet phase with
restored symmetries and no more staggered chirality. 

At last, let us comment on the  BKT algebraic phase expected at low temperature from the remnant $SO(2)$ symmetry of the model.
For most points of the phase diagram, the system shows indeed at low enough temperatures a Quasi-Long-Range-Order (QLRO)
in the $XY$ plane. There are, however, some notable exceptions to this scenario.

{\it i)} The quasicollinear pseudoplateau states

As mentioned above, there are regions in the phase diagram in which the systems shows a pseudoplateau 
at a value of the magnetisation corresponding to one third of the saturation value: in the HFP1, and in SLE1 and EL. In these regions of  parameter 
space the in-plane QLRO disappears and the BKT temperature shrinks to zero. 
The quasicollinear configurations is selected by an order by disorder mechanism due to the presence of soft modes \cite{MikeKagH}.

{\it ii)}  The classical $XY$ spin liquid point HFP2

\begin{figure}[h!]
\includegraphics[width=0.9\columnwidth]{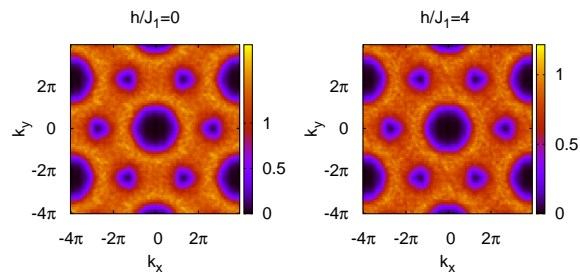}
\caption{\label{fig:SqHFP2} Structure factors for HFP2. Left: isotropic model ($h/J_1=0$). Right: components of the spins
perpendicular to the external field $h/J_1=4$. Results obtained with Monte Carlo simulations for $L=30$ lattices, at $T/J_1=2\times10^{-4}$. }
\end{figure}

In the HFP2, our simulations results seem to indicate that there is no in-plane QLRO at any value of the magnetic field.
The properties of the system at this point of the parameter space, and in the presence of a magnetic field
are remarkably similar to the ones of the isotropic case. To illustrate this, we compare the structure factor obtained for a $L=30$ system 
at $T/J_1=2\times10^{-4}$ for $h/J_1=0$ and $h/J_1=4$, shown in Fig. \ref{fig:SqHFP2}, left. For $h=0$, as commented in Ref.[\onlinecite{MoessnerkagomeHFP2}],
there are no peaks indicating that there is not thermal selection and no ``pinch-points'', consistent with a non-algebraic spin liquid phase scenario.
We show  as a typical example at $h/J_1=4$ the component perpendicular
to the magnetic field in Fig. \ref{fig:SqHFP2}, right. It shows a similar behavior to the $h=0$ case, thus suggesting that there is no QLRO. 
This result is also backed up by the value of the specific heat at low temperatures, which gives again the value of $1/2$. 

As we mentioned in the previous section, it is possible to map this HFP2 point to two models with $XXZ$ and antisymmetric DM interactions at first and second
nearest neighbors and pure exchange coupling for third nearest neighbors, models $\mathcal{H}^{HFP2}_{\chi^{\pm}}$ from Eq.(\ref{eq:hfp2xxz}).
The transformation to these models leaves the $z$ components of the spins unchanged. Therefore, the low temperature behavior of these
models under a magnetic field, $\mathcal{H}^{HFP2}_{\chi^{\pm}} + \mathcal{H}_M$, will be as we just discussed: no QLRO is expected. An interesting feature is that 
the first and second nearest neighbor DM interactions induce a non zero chirality in both first and second nearest neighbor triangles. Therefore, these
models have a net chirality with no QLRO, and we may consider them ``classical chiral spin liquids''.

\section{Dzyaloshinskii Moriya interaction} \label{sec:HDM}

In the last part of this work, we consider also the inclusion of a DM interaction. 
As mentioned when presenting the model in section I, we studied two types of nearest-neighbor DM
interaction: a uniform one (Eq.(\ref{eq:DMeq})) and a staggered one (Eq.(\ref{eq:DMdiff})).

\subsection{Homogeneous DM interaction}

The effect of an antisymmetric DM term in the nearest neighbor kagome antiferromagnet has been thoroughly studied 
\cite{CanalsLacroixDM,MikeOctu,Jaubert2017,LhuillierDM1,LhuillierDM2}.
Respecting the kagome lattice symmetries, the DM term should be perpendicular to the plane and the same in both up and down triangles.
This uniform DM is compatible with all the lattice symmetries, as discussed in section II.
It is well known that for nearest neighbor interactions in the kagome lattice this term favors the $q=0$ order with a non-zero scalar chirality.
This is also the case for the degenerate lines in the Fig.{\ref{fig:PhaseD}} phase diagram.
We thus focus on the {\it cuboc} phases, where the competition between interactions renders a richer low temperature phenomenology.

\begin{figure}[h!]
\includegraphics[width=0.94\columnwidth]{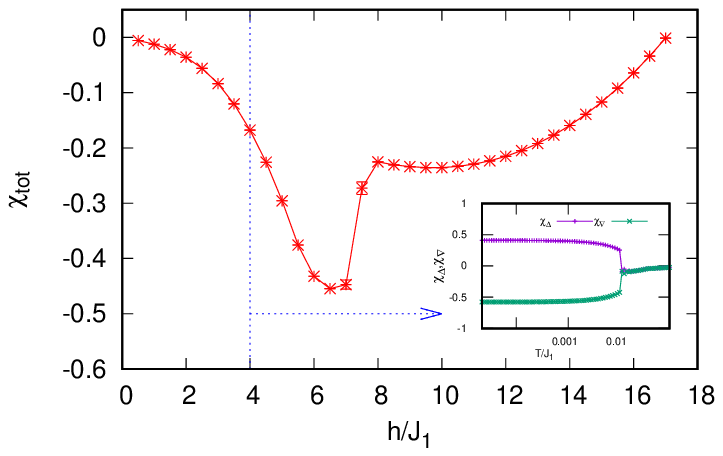}
\caption{\label{fig:ChiDMnoalt} Total chirality $\chi_{tot}$ per triangle as a function of an external 
magnetic field $h/J_1$  for $J_2=J_3=1.5J_1$, homogeneous $D=0.1J_1$ at $T/J_1=2\times 10^{-4}$,
obtained from Monte Carlo simulations for 20 copies of $L=24$. The inset shows the behavior of $\chi_{\trigup},\chi_{\trigdown}$ with temperature 
for one copy at $h/J_1=4$.}
\end{figure}

At low values of $D$, $D/J_1 \sim 0.1$, a ``cuboc-like behavior'' remains at lower magnetic fields. Along all the magnetization curve,
there is a net scalar chirality. However, this chirality is obtained in a different way depending on the coupling and the magnetic field.
For the {\it cuboc-2} phase,
at lower magnetic fields, translation symmetry is broken and 
the up and down triangles have alternating scalar chiralities with  opposite sign, that add up to a total non-zero chirality.
This is a departure from the pure {\it cuboc-2}, where there is an alternate chirality and $\chi_{\trigup}=-\chi_{\trigdown}$.
At higher fields, the $q=0$ order becomes the preferable configuration, and both chiralities acquire the same value: $\chi_{\trigup}=\chi_{\trigdown}$. In the top panel of 
Fig. \ref{fig:ChiDMnoalt}
we show simulation results for the  value of the total chirality $\chi_{tot}=\chi_{\trigup}+\chi_{\trigdown}$ 
as a function of the magnetic field at low temperature.  The inset 
shows $\chi_{\trigup}$ and $\chi_{\trigdown}$ as a function of temperature for one particular copy at $h/J_1=4$.
A similar behavior is found in the {\it cuboc-1} region. In this case, in the isotropic model the staggered chirality is
in the second-nearest neighbor triangles. Defining these chiralities  $\chi_{\rhd},\chi_{\lhd}$ in analogy to $\chi_{\trigup},\chi_{\trigdown}$ 
from Eq.(\ref{eq:Chir}), we find a net chirality at low magnetic fields for staggered $\chi_{\rhd},\chi_{\lhd}$, and a $q=0$ order at higher fields.

\subsection{Staggered DM interaction}

The staggered DM term breaks the total ({\it i.e. lattice and spin}) inversion and reflexion symmetries, but preserves the continuous $SO(2)$ spin rotation symmetry and the discrete $(x,y) \to (x, -y)$ lattice only reflection symmetry. 
In contrast to the case without DM term, no more highly degenerate points are found in the phase diagram, 
and the system shows everywhere a finite temperature BKT transition to an in-plane QLRO state. 
There are, however, different interesting phase with translation or the $(x,y) \to (x, -y)$ discrete symmetries being spontaneously broken. 

The DM term induces opposite chiralities in the up and down triangles of the kagome lattice. Therefore, in the {\it cuboc-2}
phase it favours one among the two degenerate staggered 
chirality configurations found in the isotropic case. As such, the spontaneous $\mathbb{Z}_2$ symmetry breaking for a spontaneous
staggered chirality of this case does not occur here. We have however observed a phase transition to a low temperature broken translation 
symmetry phases with striped patterns in many areas of the phase diagram: the SLE1 and SLE2 lines and the {\it cuboc-2} phase for intermediate $D/J$ values.
These broken translation phases are detected by studying the $S^z$ components of the spins. The stripes have a periodicity in the transverse direction of two or four. They 
are found in each of the site sublattices (1,2,3) of the kagome lattice, and the pattern and direction may change in each sublattice.
These site sublattices form a triangular lattice (see Fig. \ref{fig:latt}), where
the bonds are third-nearest neighbor bonds. In our model, only one bond is non-zero, the $J_3$ bond. This bond is in the $\vec{c}=-\vec{a}+\vec{b}$ direction
for sublattice 1, $\vec{b}$ direction for sublattice 2 and $\vec{a}$ for sublattice 3.
Characterising the different types of stripes found in each case in each sublattice is beyond
the scope of this work. Below, we define two types of complex order parameters, $\phi_{\vec{\alpha}}^{\gamma,\vec{\beta}},\psi_{\vec{\alpha}}^{\gamma,\vec{\beta}}$,
which can be used to study each specific case. We then illustrate their use 
in the {\it cuboc-2} phase, where the phases at higher $D/J$ are particularly interesting.

\bea 
\phi_{\vec{\alpha}}^{\gamma,\vec{\beta}} = \sum_{\beta}\sum_{\alpha} (S_i^z - S_{i+1}^z) + i( S_{i+2}^z - S_{i+3}^z) \label{eq:OP}\\
\psi_{\vec{\alpha}}^{\gamma,\vec{\beta}} = \sum_{\beta}\sum_{\alpha} (S_i^z - S_{i+2}^z) + i( S_{i+1}^z - S_{i+3}^z) \label{eq:OP22}
\eea

\noindent  where $\gamma=1,2,3$ indicates the sublattice, $S_j^z$ are four consecutive spins in each sublattice $\gamma$ summed 
along the   $\vec{\alpha}$ direction, and averaged in parallel lines in the $\vec{\beta}$ direction 
($\vec{\alpha},\vec{\beta}=\vec{a},\vec{b},\vec{c}$).

We now focus our attention in the {\it cuboc-2} area, where the low temperature phases induced by the Zeeman and antisymmetric interactions are
non trivial. The staggered DM interaction induces three types of phases, described below.

{\it i)} Low values of the DM interaction $D/J_1$ 

\begin{figure}[h!]
\includegraphics[width=0.8\columnwidth]{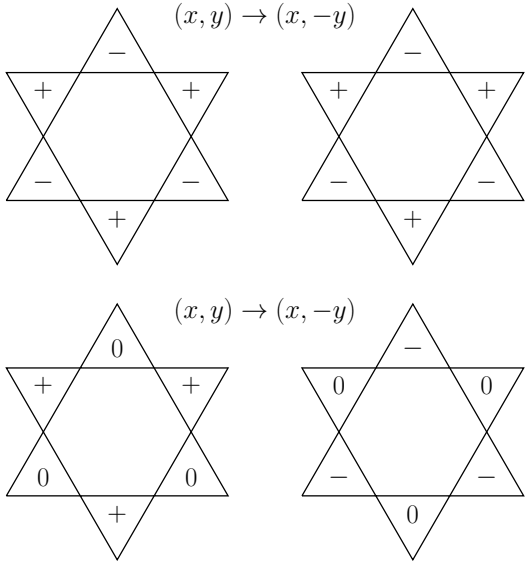}
\caption{\label{fig:RefSym} Two examples of the application of the $(x,y)\to(x,-y)$  lattice only reflection symmetry 
to the the $\chi_{\trigup,\trigdown}$ chiralities. $+,-,0$ indicate
positive, negative or zero chirality per triangle.}
\end{figure}

For lower values of $D/J_1$, at low temperatures and lower values of the magnetic field, the magnetic order
remains as the one of the   {\it cuboc-2} phase. However,  there is no more spontaneously symmetry breaking, 
since, as mentioned above, the  staggered chirality is now
fixed by the DM interaction. Note that the $(x,y)\to(x,-y)$ lattice only reflection transforms $\chi_{\trigup,\trigdown}\to-\chi_{\trigdown,\trigup}$,
and therefore a staggered chirality 
(with equal magnitude of the chirality in the up and down triangles) is invariant under this transformation, as illustrated in the top of Fig. \ref{fig:RefSym}.

{\it ii)} Intermediate $D/J_1$

\begin{figure}[h!]
\includegraphics[width=0.95\columnwidth]{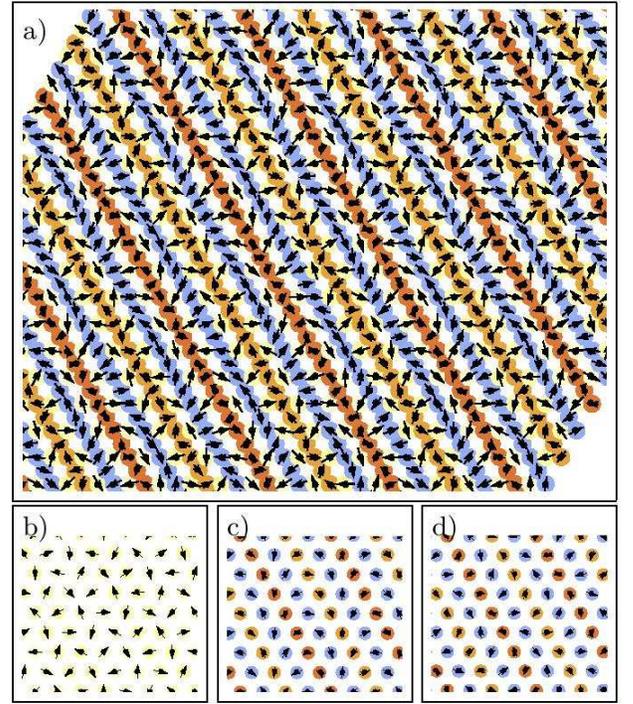} 
\caption{\label{fig:SnapDM025c2} Snapshots for the low temperature phase for $J_2=J_3=1.5J_1$, alternating DM interaction $D/J_1=0.25$, magnetic field $h/J_1=2$.
A fragment of the lattice is shown in (a), and sublattices for sites type 1,2,3 are depicted in (b),(c),(d) respectively. The colors indicate the projection along the field,
red being completely parallalel and blue antiparallel, arrows indicate the XY components. Results shown where obtained from a 
$T/J_1=2\times 10^{-4}$ simulation run in a  $L=24$ lattice.}
\end{figure}

At intermediate values of the DM parameter,
the  low temperature phase also shows a staggered chirality $\chi_{\trigup}=-\chi_{\trigdown}$ pattern,
but translation symmetry is broken in a particular way. 
The
sublattices ($1,2,3$) break translation symmetry in a striped $S^z$ pattern with a periodicity of four sites. More precisely, 
two of the sublattices show a pattern in which the components $S^z$ take three different values and are grouped $ABAC$ 
 while the third sublattice shows a pattern with two values of $S^z$ arranged in pairs $DDEE$.
To have an energetically favorable stripe ordering in the sublattices, the $ABAC$
stripes are present in two sublattices in the direction of the  $J_3$ bond of the third sublattice, which presents the $DDEE$ stripes.
For example, for stripes in the  $\vec{c}$ direction, the $ABAC$ order is in sublattices 2 and 3, and the $DDEE$ pattern  in 
sublattice 1, as shown in Fig. \ref{fig:SnapDM025c2} for $J_2=J_3=1.5J_1$,$D/J_1=0.25$,$h/J_1=2$.

\begin{figure}
\includegraphics[width=0.9\columnwidth]{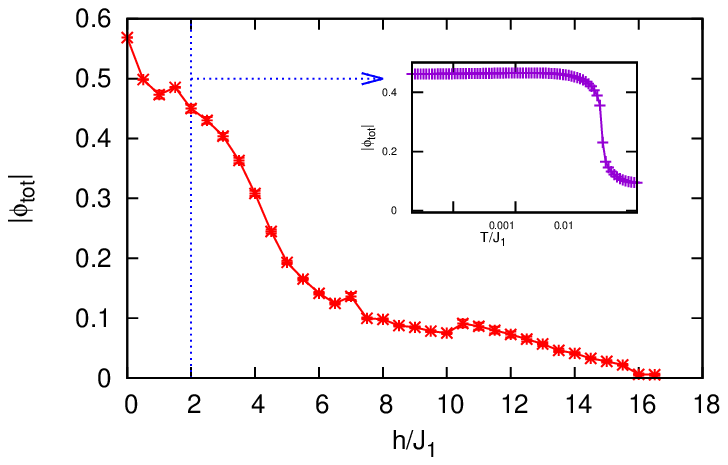}
\includegraphics[width=0.9\columnwidth]{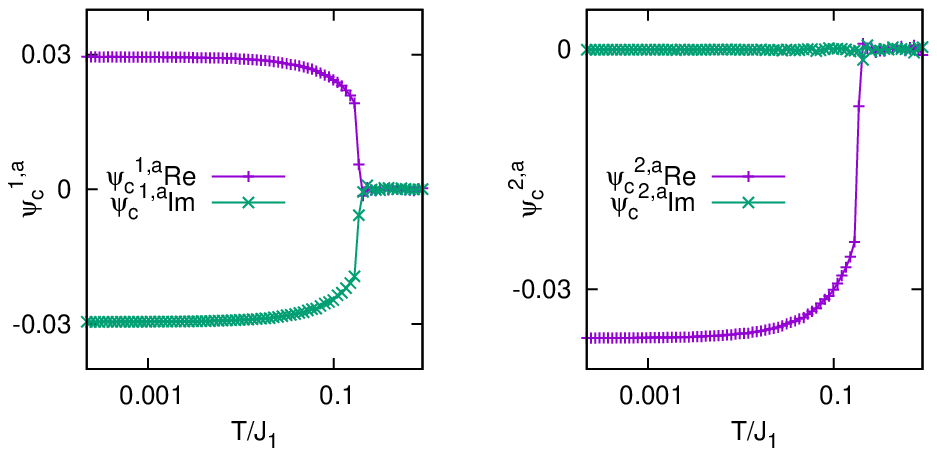} 
\caption{\label{fig:OPvHCuboc} Top: Absolute value of the total parameter $\phi_{\vec{\alpha}}^{\gamma,\vec{\beta}}$  as a function of an external 
magnetic field $h/J_1$  for $J_2=J_3=1.5J_1$, alternate $D=0.25J_1$ at $T/J_1=2\times 10^{-4}$,
obtained from Monte Carlo simulations for NNN copies of $L=24$. The inset shows the behavior of the parameter with temperature for one copy at $h/J_1=2$.
Bottom: for one copy at $h/J_1=2$, real and imaginary compoments of $\psi_{\vec{a}}^{1,\vec{c}}$ (left) and $\psi_{\vec{a}}^{2,\vec{c}}$ (right) as a function of temperature.}
\end{figure}

Let us take a moment to discuss how to stuy the order presented in the example shown in Fig. \ref{fig:SnapDM025c2} 
using the order parameters defined in Eq.(\ref{eq:OP}) and Eq.(\ref{eq:OP22}). In this case, on one hand, for 
sublattice 1 all $\phi_{\vec{\alpha}}^{1,\vec{\beta}}=0$. 
The only non zero parameters are $\phi_{\vec{a},\vec{b}}^{2,3,\vec{b},\vec{a}}$. On the other hand, 
the $\psi_{\vec{a},\vec{b}}^{\gamma,\vec{c}}$ parameters will be non-zero in all three sublattices. However, for the $DDEE$
order the real and imaginary parts of the parameter in sublattice 1,$\psi_{\vec{a},\vec{b}}^{1,\vec{c}}$, should be equal. 
For the  $ABAC$ stripes, the parameters, in this example  $\psi_{\vec{a},\vec{b}}^{2,3,\vec{c}}$, are either real or imaginary.
To illustrate this, in the top panel of Fig. \ref{fig:OPvHCuboc} we show the
absolute value at low temperature of the sum of all $\phi_{\vec{\alpha}}^{\gamma,\vec{\beta}}$ parameters as a function of the magnetic field
obtained from Monte Carlo simulations for $J_2=J_3=1.5J_1$, $D/J_1=0.25$. One can clearly see that its value decreases by increasing the magnetic field,
but remains non zero up to the saturation value 
$h/J_1 \sim 18$. The inset shows the behavior of this parameter with temperature for the case shown in  Fig. \ref{fig:SnapDM025c2}. In the lower panels, we plot
$\psi_{\vec{a}}^{1,\vec{c}}$ and $\psi_{\vec{a}}^{2,\vec{c}}$ as a function of temperature for the case of  Fig. \ref{fig:SnapDM025c2}. Since the stripes 
are in the $\vec{c}$ direction, $\text{Re}(\psi_{\vec{a}}^{1,\vec{c}})=\text{Im}(\psi_{\vec{a}}^{1,\vec{c}})$ and for this copy $\psi_{\vec{a}}^{2,\vec{c}}$ 
is real, then $\text{Im}(\psi_{\vec{a}}^{2,\vec{c}})=0$.

{\it iii)} Higher values of $D/J_1$

\begin{figure}[h!]
\includegraphics[width=0.94\columnwidth]{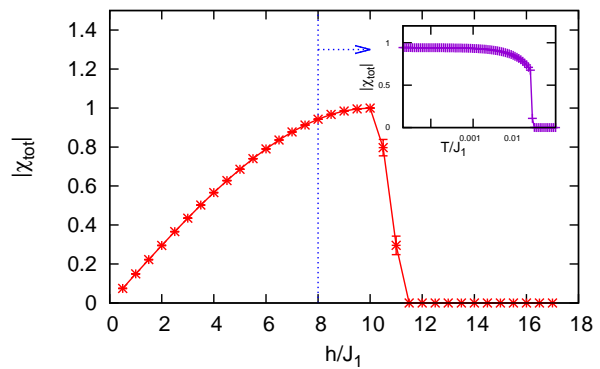}
\caption{\label{fig:ChivHCuboc} Absolute value of the total chirality $\chi_{tot}$ per triangle as a function of an external 
magnetic field $h/J_1$  for $J_2=J_3=1.5J_1$, alternate $D=0.5J_1$ at $T/J_1=2\times 10^{-4}$,
obtained from Monte Carlo simulations for 10 copies of $L=24$. The inset shows the behavior of $\chi_{tot}$ with temperature for one copy at $h/J_1=8$.}
\end{figure}

\begin{figure}[h!]
\vspace{0.5cm}

\includegraphics[width=0.95\columnwidth]{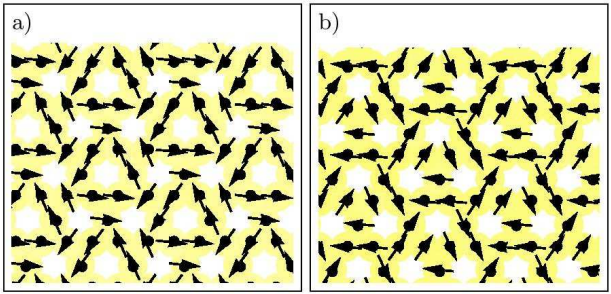}
\caption{\label{fig:SnapDM05c2} Snapshots for two different copies of the low temperature phase for $J_2=J_3=1.5J_1$, alternating DM interaction $D/J_1=0.5$,
$h/J_1=4$. a) $\chi_{\trigup}>0,\chi_{\trigdown}=0$  b) $\chi_{\trigup}=0,\chi_{\trigdown}<0$.
Results shown were obtained from 
$T/J_1=2\times 10^{-4}$ simulation runs in a  $L=30$ lattice.}
\end{figure}

 Finally, at higher values of  $\vec{D}/J_1$, the system presents an interesting
 phase with non-zero total chirality and restored translation symmetry. In this phase the lattice only reflection symmetry $(x, y) \to (x,-y)$ is spontaneously broken:
 the total chirality is a result of either $\chi_{\trigup} > 0, \chi_{\trigdown}=0$
or  $\chi_{\trigup} = 0, \chi_{\trigdown}<0$, as shown in the bottom of Fig. \ref{fig:RefSym}.
In Fig. \ref{fig:ChivHCuboc} we present the absolute value of the total chirality $|\chi_{tot}| = |\chi_{\trigup}+\chi_{\trigdown}|$ as a function
of the external magnetic field obtained from Monte Carlo simulations for $J_2=J_3=1.5J_1$, showing that this spontanesouly broken refelexion symmetry phase persists
until a critical value of the magnetic field at which the chirality vanish. The inset shows the phase transition from a paramagnetic phase to the chiral phase for a specific value of the magnetic field.
As we mentioned above, the chiral phase restores the translation symmetry, the spins 
have a uniform $S^z$ magnetization. The local order giving rise to this non-zero total chirality corresponds to, for example, up triangles 
having a non-coplanar spin configuration with non-zero chirality and  down triangle with parallel spins, and as such having zero chirality. 
In Fig. \ref{fig:SnapDM05c2} we show this QLRO in  fragments of  snapshots of the low temperature phases for different copies 
for $J_2=J_3=1.5J_1$.

\vspace{0.5cm}
Finally, in the EL, $J_1=J_3 < J_2$, the DM interaction couples the three intertissued second nearest-neighbors kagome sublattices.
Interestingly, at lower values of the magnetic field, the system shows the same broken lattice only reflection symmetry $(x, y) \to (x,-y)$ phase with total chirality
that we have just described.

\section{Discussion and conclusions}

The kagome lattice has been the laboratory for many interesting phenomena in magnetism, either at the classical and quantum level.
In this paper we have studied a classical spin system with antiferromagnetic first, second and third neighbors couplings, as well as the presence of
a magnetic field and an out of plane DM interaction. The phase diagram of the isotropic system shows lines and points of extensive or semi extensive degeneracy that keep an interesting behavior in the presence of a magnetic field. Indeed, in the lines dubbed EL and SEL1 we find a pseudo-plateau in the magnetisation curve at $1/3$ of the saturation value,
with $UUD$ and $UUUUDD$ configuration in triangle and hexagons respectively for SEL1 and $UUD$ for second neighbors triangles in EL. 
As for the first neighbors model, the BKT temperature is expected to vanish at the value of the magnetic field at which the pseudo plateau occurs. 

Of particular interest is the point dubbed HFP2: in the isotropic case it was shown to give rise to a short range spin liquid \cite{MoessnerkagomeHFP2}. 
Interestingly, this behavior seems to persist in the presence of a magnetic field, at which we claim that the BKT temperature 
is reduced to zero for the entire range of the magnetic field. Moreover, using a mapping consisting of different 
local rotations per sub-lattice  \cite{JaubertNatComm}, the model at the HFP2 point with a magnetic field can be shown to be equivalent
to a model with non-zero total
chirality an a short range classical spin liquid behavior, namely a chiral short range classical spin liquid. 

At last, we also studied the system in the combined presence of a magnetic field and a DM interaction.
More precisely we considered two cases for the DM interaction: an homogenous and a staggered DM term.
The first term is known to favour the $q=0$ configuration with non-zero total chirality.
In this sense, the interesting issue here was to study the transition from a staggered chiral configuration at the {\it cuboc-2} phases 
at low DM field to the $q=0$ chiral state at larger values of the DM term. Another interesting issue, also concerning the {\it cuboc} phases
(here we concentrated mainly in the {\it cuboc-2} phase), is the behavior of the system in the presence of a staggered DM interaction. 
At fixed values of the Heisenberg couplings and magnetic field, by increasing the DM term one passes from a {\it cuboc-2}-like 
staggered chirality phase, to a stripe phase to finally, at higher values of the DM term, a 
state with total chirality and restored translation symmetry. Moreover, the total chirality can be of the two signs, {\it i.e.} 
there is a spontaneous lattice reflexion symmetry breaking. Indeed, for those values of the parameters, 
by decreasing the temperature the system undergoes a phase transition from a normal paramagnet to a chiral state, with either positive or negative chirality.

Classical chiral states as the ones we have studied here are a good starting point in the quest of quantum chiral spin liquid,
which is very active subject in modern condensed matter physics. 
The classical states by their own are also interesting in the realisation 
of the anomalous Hall effect and  Chern insulators \cite{Ohgushi}, characterised among others by the presence of chiral edge modes. 
If the magnetic background (upon cooling) can acquire, by a spontaneous symmetry breaking, positive and negative sign for the total chirality, 
this open the possibility for tailoring an anomalous Hall effect state, 
or Chern insulator in which the edge states can go one way or the other (or the Chern number of the occupied band can be either positive or negative).
Such kind of spontaneous classical chiral states have already be seen in the context of the frustrated Heisenberg model in the triangular lattice, 
in which a skyrmion lattice configuration emerges \cite{SkyOkubo}. However, in such configurations the unit cell of the magnetic background is very large,
making the electronic transport analysis (with many bands emerging) more cumbersome.
The extended kagome model with staggered DM interaction studied here give rise to a similar behavior but with a much smaller unit cell, 
and in this sense it may be a good starting point for studying anomalous Hall effect with spontaneous broken symmetry.  

\section{ACKNOWLEDGEMENTS}

It is our pleasure to acknowledge very helpful discussion with Nathalie Guih\'ery and Mike  Zhitomirsky. F.A.G.A. thanks the Laboratoire de Physique Th\'eorique
(LPT) in Toulouse for their hospitality. This visit to LPT was financed by the Nano, EXtreme measurements $\&$ Theory  (NEXT) program  and by CONICET.
F.A.G.A. is partially supported by 
PIP 2015-0813 CONICET.


\begin{thebibliography}{99} 

\bibitem{ReviewNorman}
M. R. Norman; Rev. Mod. Phys. {\bf 88}, 041002 (2016)



\bibitem{Domenge12subl}
J.-C. Domenge, P. Sindzingre, C. Lhuillier, and L. Pierre; Phys. Rev. B {\bf 72}, 024433 (2005)
\bibitem{Richter12subl}
O. Janson, J. Richter, and H. Rosner; Phys. Rev. Lett. {\bf 101} 106403 (2008)
\bibitem{MessioCuboc}
J. -C. Domenge, C. Lhuillier, L. Messio, L. Pierre, and P. Viot
Phys. Rev. B {\bf 77}, 172413  (2008)
\bibitem{MessioPhases}
L. Messio, C. Lhuillier and G. Misguich, Phys. Rev. B {\bf 83}, 184401 (2011)

\bibitem{MessioCSL}
S. Bieri, L. Messio, B. Bernu, and C. Lhuillier, Phys. Rev. B {\bf 92}, 060407(R) (2015)
\bibitem{ShengBalents1}
S. Gong, W. Zhu, K. Yang, O. Starykh, D. N. Sheng, and L. Balents; Phys. Rev. B {\bf 94}, 035154 (2016)

\bibitem{ShengSciRep}
S. Gong, W. Zhu and D. N. Sheng, Scientific Reports {\bf 4}, 6317 (2014)
\bibitem{ShengBalents2}
S. Gong, W. Zhu, L. Balents and D. N. Sheng, Phys. Rev. B {\bf 91}, 075112 (2015)
\bibitem{MikeOctu}
A. L. Chernyshev and M. E. Zhitomirsky; Phys. Rev B  {\bf 92}, 144415 (2015)
\bibitem{CanalsLacroixDM}
M. Elhajal, B. Canals, and C. Lacroix; Phys. Rev. B {\bf 66}, 014422 (2002)
\bibitem{LhuillierDM1}
L. Messio, O. C\'epas, and C. Lhuillier;  Phys. Rev. B {\bf 81}, 064428 (2010).
\bibitem{LhuillierDM2}
L. Messio, S. Bieri, C. Lhuillier and B. Bernu; Phys. Rev. Lett. {\bf 118}, 267201 (2017)
\bibitem{Jaubert2017}
K. Essafi, O. Benton and L. D. C. Jaubert; eprint arXiv:1706.09101 (2017)
\bibitem{MC}
K. Penc, N. Shannon, H. Shiba, and Phys. Rev. Lett. {\bf 93},
197203 (2004); A. Miyata et al., J. Phys. Soc. Jpn. {\bf 80}, 074709
(2011)
\bibitem{MessioPhaseD}
L. Messio, B. Bernu, and C. Lhuillier; Phys. Rev. Lett. {\bf 108}, 207204 (2012)
\bibitem{Vish}
Itamar Kimchi and Ashvin Vishwanath
Phys. Rev. B {\bf 89}, 014414 (2014)


\bibitem{MoessnerkagomeHFP2}
J. Rehn, Arnab Sen, and R. Moessner
Phys. Rev. Lett. {\bf 118}, 047201 (2017) 

\bibitem{OBDVillain}
J. Villain et al., J. Phys. (Paris) {\bf 41}, 1263 (1980); E. F. Shender, Zh.
Eksp. Teor. Fiz. {\bf 83}, 326 (1982); C. L. Henley, Phys. Rev. Lett. {\bf 62},
2056 (1989).
\bibitem{kagomeOBD}
J. T. Chalker, P. C. W. Holdsworth, and E. F. Shender,
Phys. Rev. Lett. {\bf 68}, 855 (1992).


\bibitem{MoessnerHoneyJ3}
J. Rehn, A. Sen, K. Damle, and R. Moessner
Phys. Rev. Lett. {\bf 117}, 167201 (2016)

\bibitem{MoessnerPiro}
R. Moessner and J. T. Chalker, Phys. Rev. Lett. {\bf 80}, 2929 (1998); R. Moessner and J. T. Chalker, Phys. Rev. B {\bf 58}, 12049 (1998).









\bibitem{JaubertNatComm}
K. Essafi, O. Benton and L. D. C. Jaubert; Nature Communications {\bf 7},10297 (2016)
\bibitem{MikeKagH}
M. E. Zhitomirsky,  Phys. Rev. Lett. {\bf 88}, 057204 (2002); M V Gvozdikova1, P-E Melchy and M E Zhitomirsky
Journal of Physics: Condensed Matter {\bf 23} 16 (2011) 


\bibitem{Ohgushi}
K. Ohgushi, S. Murakami, and N. Nagaosa
Phys. Rev. B {\bf 62}, R6065 (2000)

\bibitem{SkyOkubo}
 T. Okubo, S. Chung, and H. Kawamura; Phys. Rev. Lett. {\bf 108}, 017206 (2012)

\end{thebibliography}
\end{document}